\documentclass[backend=biber]{gistt}

\usepackage{blindtext}
\usepackage{soul}
\usepackage{soulpos}
\usepackage[x11names]{xcolor}
\usepackage{tikz}
\usetikzlibrary{arrows.meta}
\usepackage{nicefrac}
\usepackage{listings}
\usepackage{lstautogobble}
\usepackage{todonotes}

\DeclareSourcemap{
  \maps[datatype=bibtex, overwrite]{
    \map{
      \step[fieldset=note, null]
      \step[fieldset=file, null]
      \step[fieldset=keywords, null]
      \step[fieldset=language, null]
      \step[fieldset=editor, null]
      \step[fieldset=annotation, null]
      \step[fieldset=abstract, null]
      \step[fieldset=location, null]
      \step[fieldset=urldate, null]
      \step[fieldset=url, null]
      \step[fieldset=publisher, null]
      \step[fieldset=booktitle, null]
      \step[fieldset=pages, null]
      \step[fieldset=issn, null]
      \step[fieldset=isbn, null]
      \step[fieldset=eventtitle, null]
      \step[fieldset=institution, null]
      \step[fieldset=eprint, null]
      \step[fieldset=eprinttype, null]
      \step[fieldset=eprintclass, null]
      \step[fieldset=edition, null]
      \step[fieldset=pagetotal, null]
    }
    \map[overwrite]{
      \step[fieldsource=shortjournal]
      \step[fieldset=journaltitle,origfieldval]
    }
  }
}

\bibliography{references.bib}

\definecolor{ACMRed}{rgb}{.992,.106,.078}
\definecolor{ACMOrange}{rgb}{.988,.573,0}
\definecolor{ACMPurple}{rgb}{.396,.004,.42}
\definecolor{ACMBlue}{rgb}{.004,0.51,.675}
\definecolor{ACMDarkBlue}{rgb}{.035,.208,.478}

\ulposdef{\bluehighlight}{%
  \mbox{\color{LightSteelBlue1}\rule[-.8ex]{\ulwidth}{13pt}}%
}
\newcounter{tipcounter}
\NewDocumentCommand{\tip}{}{%
  \addtocounter{tipcounter}{1}%
  \bluehighlight{\:Recommendation~\thetipcounter\:}%
}

\title{Benchmarking Function Hook Latency in Cloud-Native Environments}

\author{%
\href{https://orcid.org/0000-0002-6820-4953}{Mario Kahlhofer}\textsuperscript{1},
\href{https://orcid.org/0009-0006-6932-3045}{Patrick Kern}\textsuperscript{1},
\href{https://orcid.org/0000-0001-6912-2549}{Sören Henning}\textsuperscript{1,2},
\href{https://orcid.org/0000-0003-2821-2489}{Stefan Rass}\textsuperscript{2}\\
\{mario.kahlhofer, patrick.kern\}@dynatrace.com,
\{soeren.henning, stefan.rass\}@jku.at\\
\textsuperscript{1}Dynatrace Research,
\textsuperscript{2}Johannes Kepler University Linz}

\newcommand{\TestTotalSampleSize}{50,000}
\newcommand{\TestWarmUpPeriodSize}{4,000}
\newcommand{\TestMeasurementPeriodSize}{46,000}

\newcommand{\TestResultLocalStdPool}{0.50}

\newcommand{\TestResultKindStdPool}{0.84}

\newcommand{\TestResultSamePodStdPool}{1.64}
\newcommand{\TestResultSameClusterPValue}{0.2713}
\newcommand{\TestResultSameClusterStdPool}{1.95}
\newcommand{\TestResultDockerSameClusterStdPoolRate}{3.9}

\begin{document}

\maketitle

\begin{abstract}
  % State the problem
  Researchers and engineers are increasingly adopting cloud-native technologies
  for application development and performance evaluation.
  While this has improved the reproducibility of benchmarks in the cloud,
  the complexity of cloud-native environments makes it difficult to run benchmarks reliably.
  Cloud-native applications are often instrumented or altered at runtime,
  by dynamically patching or hooking them, which introduces a significant performance overhead.
  % Say what your solution achieves
  Our work discusses the benchmarking-related pitfalls of the dominant
  cloud-native technology, Kubernetes, and how they affect
  performance measurements of dynamically patched or hooked applications.
  % Say what follows from you solution
  We present recommendations to mitigate these risks
  and demonstrate how an improper experimental setup can
  negatively impact latency measurements.
\end{abstract}

\section{Introduction}

Cloud-native technologies aim to build
loosely coupled, resilient, observable, and secure systems%
~\cite{CNCF2018:CloudNativeDefinition}.
Observability and security are typically achieved by
dynamically instrumenting or altering already built applications
with \emph{function hooks}~\cite{Lopez2017:SurveyFunctionSystem}.
These are small pieces of code added to an application's functions.
In particular, security tools need to dynamically
modify, redirect, or block specific execution patterns,
which often results in significant performance penalties%
~\cite{Viktorsson2020:SecurityPerformanceTradeoffsKubernetes}.

Careful benchmarking is required to measure the performance impact of such changes.
Besides \emph{empirical standards} for software benchmarking%
~\cite{Ralph2021:EmpiricalStandardsSoftware}
and \emph{methodological principles} for performance evaluation in cloud computing%
~\cite{Papadopoulos2019:MethodologicalPrinciplesReproducible},
we address benchmarking-related pitfalls of cloud-native environments with:

\begin{enumerate}
  \item Recommendations on how to measure the latency
        of function hooks in cloud-native environments.
  \item A demonstration of an improper experimental setup
        that makes hypothesis testing harder.
\end{enumerate}

\section{Cloud-Native Benchmark Suite}

\begin{figure}[h]
  \centering
  \resizebox{0.9\columnwidth}{!}{\begin{tikzpicture}[scale=1.0]

  \node[draw, rectangle, minimum size=0.5cm, inner xsep=2mm, anchor=east, text width=2.6cm, align=center] (a) at (0,0.4) {Load Generator};
  \node[draw, rectangle, minimum size=0.5cm, inner xsep=2mm, anchor=east, text width=2.6cm, align=center] (b) at (0,-0.4) {Monitoring Tool};
  \node[draw, draw=SteelBlue3, fill=LightSteelBlue1!30, rectangle, minimum size=0.75cm, inner xsep=2mm, anchor=west, text width=3cm, align=center] (c) at (1.5,0) {System Under Test\\(SUT)};
  \draw[-{Latex[length=2.5mm]}] (a) -- (c);
  \draw[-,dashed] (b) -- (c);
  \draw[-,dotted] (a) -- (b);

\end{tikzpicture}}
  \caption{Typical components of a benchmark suite}
  \label{fig:suite}
\end{figure}
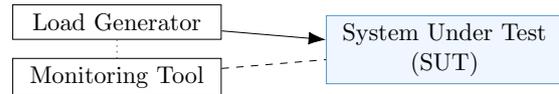

Cloud environments are frequently used to build complete benchmark suites,
as they provide a well-reproducible environment%
~\cite{Henning2021:ReproducibleBenchmarkingCloudNative}.
A typical benchmark suite (Figure~\ref{fig:suite})
consists of a \emph{system under test (SUT)}, e.g., the patched application,
a \emph{load generator} sending requests to that application,
and a \emph{monitoring tool} measuring performance metrics.
Latency is often measured directly by the load generator.

In Kubernetes, workloads are organized into \emph{pods} of one or more \emph{containers}
which share storage and networking resources. Physical or virtual machines
that run these pods are called \emph{nodes}.

\tip{}~%
When measuring latency,
ensure that the load generator and the SUT are in
separate containers within the same pod.
Otherwise, additional network hops may distort the measurements.
% to avoid a systematic error due to unpredictable network latencies.

\tip{}~%
If components of the benchmark suite need to be in separate pods,
ensure that both pods are deployed on the same physical node, e.g.,
by specifying \emph{node restrictions} in Kubernetes.

\tip{}~%
Weigh the benefits of a \emph{service mesh} against its additional network overhead.
Service meshes wrap each application behind a reverse proxy
and make it easier to monitor and control inbound and outbound network traffic%
~\cite{Henning2022:CloudNativeScalabilityBenchmarking}.

\tip{}~%
Generally avoid benchmarking in \emph{multi-tenancy clusters}, i.e.,
clusters that are shared across teams, either physically or virtually.

\section{Function Hook Granularity}

We distinguish four layers%
~\cite{Islam2023:RuntimeSoftwarePatching,Lopez2017:SurveyFunctionSystem}
where function hooks or patches can be injected (Figure~\ref{fig:stack}):

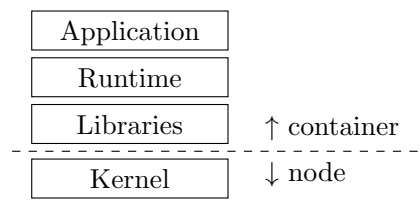
\begin{figure}[b]
  \centering
  \resizebox{0.68\columnwidth}{!}{\begin{tikzpicture}[scale=1.0]

  % inner rectangles
  \foreach \i/\title/\offset in {1/Application/0,2/Runtime/0,3/Libraries/0,4/Kernel/0.1}
    {
      \draw (0.75,4-\i*0.6-\offset) rectangle (3.25,4-\i*0.6-0.5-\offset);
      \node[anchor=mid,align=center] at (2,4-\i*0.6-0.3-\offset) {\title};
    }

  % dotted line
  \draw[dashed] (0.5,1.6) -- (5.75,1.6);

  % container and node text
  \node[anchor=west,align=left] at (3.6, 4-3*0.6-0.3) {$\uparrow$~container};
  \node[anchor=west,align=left] at (3.6, 4-4*0.6-0.3) {$\downarrow$~node};

\end{tikzpicture}}
  \caption{Typical layers of a software application}
  \label{fig:stack}
\end{figure}

\begin{itemize}
  \item \textbf{Application-level} hooks use methods
        implemented by the application's developers,
        e.g., a plugin system.
        Since such systems are not widely available, this layer cannot be used
        for general-purpose hooks on already built applications.
  \item \textbf{Runtime-level} hooks use native capabilities
        of language runtimes to modify applications, e.g.,
        the JVM Tool Interface (JVM TI),
        %\footnote{\url{https://docs.oracle.com/javase/8/docs/technotes/guides/jvmti/}},
        the .NET Profiling API,
        %\footnote{\url{https://learn.microsoft.com/en-us/dotnet/framework/unmanaged-api/profiling/profiling-overview}},
        or Node.js module preloading.
        %\footnote{\url{https://nodejs.org/api/cli.html#-r---require-module}}.
        %
  \item \textbf{Library-level} hooks override symbols
        in shared libraries, e.g., by the ``LD\_PRELOAD trick''%
        ~\cite{Lopez2017:SurveyFunctionSystem}.%
        \footnote{\url{https://man7.org/linux/man-pages/man8/ld.so.8.html}}
  \item \textbf{Kernel-level} hooks use native capabilities
        of the operating system to modify application behavior, e.g.,
        kernel modules or eBPF programs.
\end{itemize}

\tip{}~%
The monitoring tool should be placed as close as possible
to the layer where the hook is injected. Testing farther away pollutes
measurements with noise from other layers (Section~\ref{sec:results}).

To achieve optimal results, hooking and monitoring should be done at the ``same layer'', i.e.,
by embedding monitoring functionality into the hook itself, e.g.,
by recording timestamps before and after the hook is executed, directly in the hook's code.

Moving the monitoring tool further away from the hook is justifiable
when one wants to more accurately represent real-world behavior instead.

\tip{}~%
Describe if the benchmark measures the specific hooking overhead in isolation
(\emph{micro benchmark}), or rather represents a real-world application with
a hook injected into it (\emph{macro benchmark})%
~\cite{Waller2014:PerformanceBenchmarkingApplication}.
Benchmarking both cases and discussing the differences is recommended.

\section{Demonstration}

We first build a Java application that simply responds with ``Hello World'' to any HTTP request.
We then implement a \emph{library-level LD\_PRELOAD hook}
that blocks all requests that contain specific keywords (Listing~\ref{lst:hook}).
The hook changes the \texttt{read}%
\footnote{\url{https://man7.org/linux/man-pages/man2/read.2.html}}
system call
by overriding the corresponding symbol in the C standard library.%
\footnote{%
  The full source code can be found at
  \url{https://github.com/dynatrace-research/function-hook-latency-benchmarking}}
We then measure our application's network performance with and without the hook.

\noindent
\begin{minipage}{\columnwidth}
\begin{lstlisting}[
    label=lst:hook,
    firstline=15,lastline=24,
    caption={A hook on the \texttt{read} symbol of glibc}]
ssize_t read(int fd, void *buf, size_t count) {
  read_t read_ptr = (read_t)dlsym(RTLD_NEXT, "read");
  ssize_t bytes_read = read_ptr(fd, buf, count);
  if (is_http_socket(fd)) {
    if (contains_keyword(buf, count)) {
      // trace or block call
    }
  }
  return bytes_read;
}
\end{lstlisting}
\end{minipage}

Low-level function hooks carry the risk that
the hooked function is used by high-level functionality
for a purpose other than the one originally intended.
For example, the \texttt{read} call that we override here
is also used to read regular files, not just network packets.

\tip{}~%
Therefore, describe how the hooked function is typically used by applications
and ensure that the benchmarks reflect their proper use, e.g.,
with \emph{synthetic micro benchmarks}~\cite{Waller2014:PerformanceBenchmarkingApplication},
but also real-world behavior.
Suitable cloud-native, real-world reference applications are
TeaStore~\cite{vonKistowski2018:TeaStoreMicroServiceReference},
DeathStarBench~\cite{Gan2019:OpenSourceBenchmarkSuite},
or Unguard~\cite{DynatraceLLC2023:UnguardInsecureCloudnative}
for security use cases.

\subsection{Experimental Setup}

Our experiment consists of two containers: Locust
(a performance testing tool)
% \footnote{\url{https://locust.io}}
as the load generator with embedded monitoring, and the SUT.
With containers, we not only represent cloud-native paradigms,
but also isolate concerns between the \emph{benchmark owner} and the \emph{SUT owner}.
%
% Note that this violates Recommendation 5:
% Ideally, we should hook, test, and monitor within a single container.
% However, this means that we will not have the advantages of well-isolated containers.
%
We compare four conditions:

\begin{enumerate}
  \item \textbf{In Docker} (a popular container runtime):
        Both containers run on a single server, communicating through the host network.
  \item \textbf{In Kind} (a tool for running Kubernetes using Docker containers):
        Both containers run inside a single pod on a local, single-node Kind cluster.%
        % \footnote{\url{https://kind.sigs.k8s.io}}
\end{enumerate}

The remaining two conditions use the AWS EKS service
(a popular enterprise-grade cloud provider).%
% \footnote{\url{https://aws.amazon.com/eks}}

\begin{enumerate}
  \setcounter{enumi}{2}
  \item \textbf{In EKS pod}:
        Both containers run inside a single pod in a managed, single-node AWS EKS cluster.
  \item \textbf{Across EKS nodes}:
        Both containers run in separate pods, each pod on a different node,
        in a managed AWS EKS cluster with two nodes.
\end{enumerate}

Docker and Kind are running on a
\mbox{24-core} (Intel Xeon E5-2680~v3) Ubuntu~22.04 server with 64~GB memory.
EKS nodes are t3.medium EC2 instances
(2~vCPUs, Intel Xeon Platinum~8000, 4~GB memory).

\tip{}~%
Ensure that the servers do not hit any resource limits during the experiment
to avoid performance degradations due to resource contention.

\begin{figure}[t]
  \centering
  \includegraphics[width=\columnwidth,page=1]{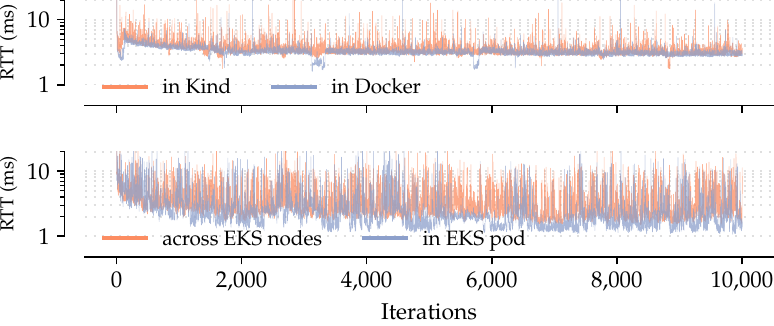}
  \caption{Logarithmic lag plot on measured RTT}
  \label{fig:lagplots}
\end{figure}

We measure the round-trip time (RTT) of $\TestTotalSampleSize{}$
HTTP request-response interchanges between Locust and the SUT.
We empirically observed that the RTT definitely stabilizes under all four conditions
after ${\sim\TestWarmUpPeriodSize{}}$ warm-up requests (Figure~\ref{fig:lagplots}).

\subsection{Hypothesis Testing and Results}
\label{sec:results}

\begin{figure}[t]
  \centering
  \includegraphics[width=\columnwidth,page=1]{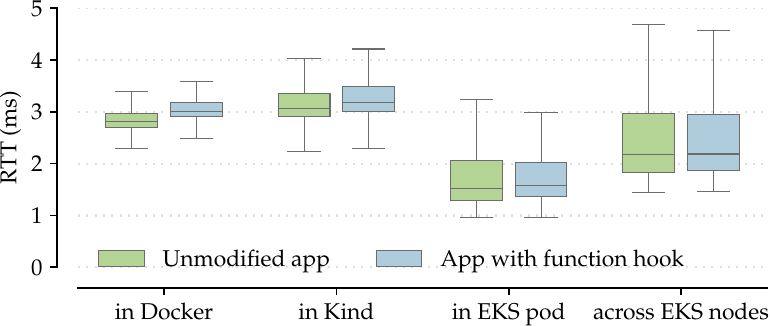}
  \caption{Tukey boxplots on \TestTotalSampleSize{} RTT measurements per condition,
    without any warm-up requests}
  \label{fig:boxplots}
\end{figure}

Figure~\ref{fig:boxplots} shows the RTT distribution per condition,
with and without the hook, after warm-up requests.

Our function hook must introduce a performance overhead:
To test the null hypothesis that the mean RTT is the same
with and without the hook, we use an independent two-sample $t$-test,
assuming equal but unknown variances and equal sample sizes.%
\footnote{We use the \texttt{stats.ttest\_ind} test from the SciPy package.}
Let $\bar{x}_1$ and $\bar{x}_2$ be the sample means,
$n$ be the sample size per condition,
and $s_p$ the \emph{pooled standard deviation}%
\footnote{
  With sample variances $s_1^2$ and $s_2^2$
  and equal sample sizes,
  the pooled variance is defined by
  ${ s_p^2 = (s_1^2 + s_2^2) \, / \, 2 }$.
}, then the test statistic is given by:
$t = (\bar{x}_1 - \bar{x}_2) \, / \, (s_p \sqrt{\nicefrac{2}{n}})$.

With $n=\TestMeasurementPeriodSize{}$~samples left per condition
after removing warm-up requests, the hooking overhead is significant (${\alpha=0.05}$)
in three of four conditions (all~${p<0.0001}$).
The \emph{across EKS nodes} condition is not significant (${p=\TestResultSameClusterPValue{}}$)
since we lose a lot of statistical power in EKS due to the higher RTT variance.

As expected and in line with related work%
~\cite{Papadopoulos2019:MethodologicalPrinciplesReproducible},
measurements taken in EKS generally exhibit
a much higher variance than measurements on our own server,
due to diverse factors that can hardly be controlled for.
We expected the \emph{Docker} condition, being the most minimal setup,
to show the lowest variance (${s_p=\TestResultLocalStdPool{}}$),
and the \emph{Kind} condition, which just adds a few Kubernetes components,
to show the second-lowest variance (${s_p=\TestResultKindStdPool{}}$).
We also expected that the network latency \emph{across EKS nodes}
shows the highest variance (${s_p=\TestResultSameClusterStdPool{}}$;
\TestResultDockerSameClusterStdPoolRate{} times higher than in \emph{Docker}).
Packets in that condition traverse the origin container, pod, and node,
some intermediate network that interconnects nodes,
until they reach the target node, pod, and container again.
In a multi-cluster environment, a communication path with that many hops is common.
Kubernetes \emph{services},
%\footnote{\url{https://kubernetes.io/docs/concepts/services-networking/service/}},
which are widely used abstractions on top of pods, would add even more hops to that route.
Perhaps surprising is that the variance of the \emph{EKS pod} condition
was still relatively high (${s_p=\TestResultSamePodStdPool{}}$).
Keeping network communication within the same pod decreased the variance,
but it seems that the background noise of our EKS cluster is still relatively high
and affecting inter-pod traffic.

\tip{}~%
As shown, measurements in cloud-native environments
tend to have a higher variance than in local environments%
~\cite{Papadopoulos2019:MethodologicalPrinciplesReproducible}.
To regain statistical power, the sample size must be increased.


% \begin{figure*}[tb!]
%   \centering
%   \includegraphics[width=0.75\textwidth]{./figures/network.png}
%   % https://excalidraw.com/#json=IGsucqu2i1L5NOzXI3qGb,-lyvqb2agw6lPEFmGuPB-Q
%   % https://sookocheff.com/post/kubernetes/understanding-kubernetes-networking-model/
%   \caption{%
%     Kubernetes network architecture,
%     where the test suite and the SUT are located in different nodes
%   }
%   \label{fig:network}
% \end{figure*}

\tip{}~%
Conducting experiments in differently configured environments
is a general principle~\cite[P2]{Papadopoulos2019:MethodologicalPrinciplesReproducible}
that is especially relevant for cloud-native environments.
Different cloud providers, service meshes, or network setups help increase diversity.

\section{Conclusion}

This work provides \thetipcounter~practical recommendations for researchers and engineers
who benchmark function hook latency in cloud-native environments,
but want to reduce the measurement bias introduced by these environments.
We have shown that function hook latency measurements can be easily contaminated by noise,
without doing anything obviously wrong.
%
% Our recommendations will not eliminate measurement bias,
We hope to raise awareness while providing practical guidance
for similar latency-based benchmarks, as some of our recommendations are also broadly applicable.

{\sloppy\hbadness=10000\printbibliography}

\end{document}